\documentclass[sigconf,authorversion,nonacm]{acmart}

\AtBeginDocument{%
  \providecommand\BibTeX{{%
    \normalfont B\kern-0.5em{\scshape i\kern-0.25em b}\kern-0.8em\TeX}}}

\begin{document}

\title{Respect for Human Autonomy in Recommender Systems}

\author{Lav R.\ Varshney$^{\dagger,\ddagger}$}
\email{varshney@illinois.edu}
\affiliation{%
  \institution{$^{\dagger}$Salesforce Research \\ $^{\ddagger}$University of Illinois at Urbana-Champaign}
}

\renewcommand{\shortauthors}{Varshney}

\begin{abstract}
Recommender systems can influence human behavior in significant ways, in some cases making people more machine-like.  In this sense, recommender systems may be deleterious to notions of human autonomy.  Many ethical systems point to respect for human autonomy as a key principle arising from human rights considerations, and several emerging frameworks for AI include this principle.  Yet, no specific formalization has been defined.  Separately, self-determination theory shows that autonomy is an innate psychological need for people, and moreover has a significant body of experimental work that formalizes and measures level of human autonomy.  In this position paper, we argue that there is a need to specifically operationalize respect for human autonomy in the context of recommender systems.  
Moreover, that such an operational definition can be developed based on well-established approaches from experimental psychology, which can then be used to design future recommender systems that respect human autonomy.     
\end{abstract}


\maketitle

\section{Introduction: Respect for Human Autonomy as an Ethical Principle}
It has oft been  recognized that recommender systems have a significant impact on human behavior in a kind of positive feedback loop \cite{KnijnenburgSW2016}---in fact in many settings such as commercial ones, this is their primary design goal \cite{Seaver2019,MilanoTF2020}.  As Wu describes: ``A platform like Facebook\dots provide[s] content that maximizes `engagement,' which is information tailored to the interests of each user. While this sounds relatively innocuous (giving users what they want), it has the secondary effect of exercising strong control over what the listener is exposed to'' \cite{Wu2018}.  These impacts on human attention and behavior may be perceived in different manners at the individual level and at the population level: although recommender systems may push individuals into considering novel possibilities, such novelty may be identical across the population and therefore cause greater convergence \cite{FlederH2009}.  These short-term disparate perceptions of impact on human behavior may be important in determining their long-term impact on human well-being.

Recent research with predictive text systems based on deep learning (such as large-scale language models) rather than, say, collaborative filtering, has experimentally demonstrated that when phones/computers suggest words or phrases in text messages and email, it changes the way people write. In particular, the writing of users of such technologies becomes more succinct, more predictable, and less colorful \cite{ArnoldCG2020}.  As such, one can observe that artificial intelligence (AI) technologies such as recommender systems that are designed to address the problem of information overload by improving the efficiency of human work may also frequently impact the content of human work in potentially unanticipated ways.

Given the strong impact that recommender systems have on human behavior, one may view them as causing a loss of human autonomy \cite{MilanoTF2020}.  
Virtually all theories of autonomy agree that two conditions are essential for autonomy: liberty (independence from controlling influences) and agency (capacity for intentional action) \cite{BeauchampC2019}.  Evidently, recommender systems deprive human users of liberty due to their controlling influences, and also often agency since human users do not usually provide informed consent when using recommender systems (users often lack the choice and are given a `take it or leave it' option when accessing online services \cite{Nissenbaum2011, TaddeoF2016}).

Respect for human autonomy is a longstanding and keystone ethical principle in biomedical ethics, alongside principles of beneficence, non-maleficence, and justice. As Beauchamp and Childress say in their treatise on the subject: ``The principle of respect for the autonomous choices of persons runs as deep in morality as any principle'' \cite{BeauchampC2019}. We have previously suggested it is also important for technology that addresses information overload \cite{Varshney2014}.  Many recent discussions of AI ethics frameworks since our prior work have also suggested respect for human autonomy as a key ethical principle.  As a typical example focusing on the negative obligations of the principle, Cowls et al. say that ''It is essential that software intervenes in users' life only in ways that respects their autonomy. Again, this is not a problem that arises only with AI-driven interventions, but the use of AI introduces new considerations'' \cite{CowlsKTF2019}.  A brief discussion specifically focused on recommender systems is given by Milano et al.\ \cite{MilanoTF2020}.  Several recent codes of conduct for AI research and development also discuss respect for human autonomy at a high level.\footnote{
For example the \emph{AI Ethics Principles} developed by the 
Department of Industry, Science,
Energy and Resources in the  Australian Government states that ``Throughout their lifecycle, AI systems should respect human rights, diversity, and the autonomy of individuals. This principle aims to ensure that AI systems are aligned with human values. Machines should serve humans, and not the other way around. \dots All people interacting with AI systems should be able to keep full and effective control over themselves.''
}  

In addition to the negative obligation that autonomous obligations not be subject to control by others, there is also a positive obligation that requires respectful disclosure of information to foster autonomous decision making \cite{BeauchampC2019}.  Physicians not disclosing to patients their use of AI recommendations for treatment options \cite{Cohen2020} is violative of the positive obligation.

One important challenge in analyzing the ethical question of human autonomy in recommender systems is that their benefits are intertwined with autonomy. As Whittlestone et al. argue \cite{WhittlestoneNAC2019}: ``Automated solutions may
genuinely improve people's lives by saving them time on
mundane tasks that could be better spent on more rewarding
activities. But they also risk disrupting some of the practices 
that are an important part of what makes us human. \dots we may also see widespread deskilling,
atrophy, ossification of practices, homogenisation and
cultural diversity.''  On a more technically-oriented point raised by \cite{CowlsKTF2019}, to obtain information that effectively personalizes recommendation, ``an intervention strategy that has no impact on user autonomy (e.g., one that lacks any interventions) may be ineffective in extracting the necessary information for correctly contextualised future interventions.'' 

Given the importance of respecting human autonomy as an ethical principle for AI systems, and the fact recommender systems may be violative, there is a surprising dearth of formalizations of autonomy that can be specifically measured in recommender systems.  This is in contrast to ethical properties of AI systems such as fairness, safety, or privacy that are starting to have well-developed mathematical formulations that yield standardized measurement approaches, e.g.\ \cite{Bellamy_ea2019}.  

Separate from ethical frameworks around human autonomy that largely stem from human rights perspectives, a branch of psychology dealing with human well-being called \emph{self-determination theory} \cite{DeciR2012} also places human autonomy at the center.  Interestingly, there are well-established operationalizations of human autonomy that are used in numerous experiments in that field.  In this position paper, we argue that there is a need to specifically operationalize respect for human autonomy in the context of recommender systems. Moreover, that such an operational definition can be developed based on well-established approaches from experimental self-determination theory, which can in turn be used to design future recommender systems that respect human autonomy.

\section{Human Autonomy in Self-Determination Theory}

Self-determination theory within positive psychology posits that human well-being thrives when three universal psychological needs are satisfied: autonomy, relatedness, and competence \cite{RyanD2000, DeciR2012}.  This in turn leads to greater creativity, effective problem solving, motivation, goal pursuit, performance, persistence, mental health, and high-quality personal relations with others.
In particular, when these three needs are satisfied, people experience events, jobs, and tasks as self-determined, and their behavior becomes intrinsically motivated from personal goals and values rather than extrinsically motivated by external reinforcements or demands. According to self-determination theory, \emph{human autonomy} is defined as having flexibility and control over processes and outcomes \cite{RyanD2006}.

Importantly, self-determination theory does not consider the three psychological needs to be equal in the experience of self-determination.  Autonomy is given the paramount role \cite{DeciR2012}.  (This is in contrast to ethical frameworks where respect for human autonomy is explicitly not given priority over all other moral principles including beneficence, non-maleficence, and justice \cite{BeauchampC2019}.) Relatedness and competence are conceptualized as part of the contextual background that gives rise to self-determination, whereas autonomy is viewed as the key ingredient. As Deci and Ryan detail: “autonomy occupies a unique position in the set of three needs: being able to satisfy the needs of competence and relatedness may be enough for controlled behavior, but being able to satisfy the need for autonomy is essential for the goal-directed behavior to be self-determined and for many of the optimal outcomes associated with
self-determination to accrue.'' 

Self-determination theory findings largely hold across cultural contexts \cite{Church_ea2013}.

Given the importance of autonomy in self-determination theory, it has not just remained as a principle or theoretical construct.  It has become an experimentally measurable quantity.  As detailed in the psychometrics and experimental psychology literature, one can develop robust tests for human autonomy.  As a typical example, one can consider the Index of Autonomous Functioning (IAF), which measures  trait autonomy based on three theoretically-derived subscales  that focus on self-congruence, interest-taking, and low susceptibility to control, respectively \cite{WeinsteinPR2012}.  

The IAF has been used, e.g.\ for assessing the impact on human autonomy of new technologies (not necessarily information technologies) for older adults \cite{KohlbacherH2016}.  In this usage, by aggregating over a population of human users, the scale is converted from measuring an individual human trait into measuring a technological system's ability to respect autonomous function. This population-level assessment enables psychometric evaluation technologies for respecting human autonomy.

This population-level psychometric testing usage of the IAF to evaluate technology can be directly adapted specifically to measure how well recommender systems respect human autonomy.  

\section{Operationalizing Human Autonomy for Recommender Systems}

In designing recommender systems that take not just relevance and predictive performance into account \cite{KnijnenburgSW2016}, but also a broader set of human values, it is important to include an appropriate set of these values into the design objectives/constraints, either through direct specification or through data-driven learning \cite{StrayAH2020}.  In thinking about including various ethical principles, it is useful to define terms unambiguously: what is meant by \emph{fairness},  \emph{autonomy}, and other such terms, and how might these be interpreted across contexts \cite{WhittlestoneNAC2019}.  Moreover, how can these properties be measured and incorporated into system design.

Here we have argued that well-established theoretical constructs and experimental psychometric methods from self-determination theory are appropriate for such operationalization.   With such an operational definition in hand, one can design recommender systems using techniques such as reinforcement learning with human feedback, which have recently been demonstrated for language model tasks \cite{Stiennon_ea2020}. 

One may wonder, however, if it is possible to operationalize the principle of respect for human autonomy in an abstract mathematical manner that avoids the need for tests with human subjects as part of a feedback-based design methodology.  This is an important open question that we aim to address in future research.

\begin{acks}
Discussions with Kathy Baxter (Salesforce Research) and Jonathan Stray (Partnership on AI) on this and related topics are gratefully appreciated.  Suggestions from the anonymous reviewers are also much appreciated.
\end{acks}

\bibliographystyle{ACM-Reference-Format}
\bibliography{sample-base}

\end{document}